\documentclass{appolb}
\usepackage{graphicx}
\usepackage{slashed}
\usepackage{float}
\usepackage{amsmath}
\usepackage{bm}
\usepackage[numbers,sort&compress]{natbib}
\begin{document}

	\title{Heavy and heavy-light tensor and axial-tensor mesons in the Covariant Spectator Theory %
		\thanks{Presented at Excited QCD Workshop 2026, Granada, Spain}%
	}
	\author{Elmar P. Biernat
		\address{Centro de Ci\^encias e Tecnologias Nucleares and Departamento de Engenharia e Ci\^encias Nucleares, Instituto Superior T\'ecnico, Universidade de Lisboa, Campus Tecnol\'ogico e Nuclear, 2695-066 Bobadela, Portugal\\
		elmar.biernat@tecnico.ulisboa.pt}
		\\[2.3mm]
		Alfred Stadler 
		\address{Departamento de F\'isica, Universidade de \'Evora, 7000-671 \'Evora, Portugal\\
	Laborat\'orio de Instrumenta\c{c}\~ao e F\'isica Experimental de Part\'iculas---LIP, Avenida Professor Gama Pinto, 2, 1649-003 Lisboa, Portugal}
	}
	
	\maketitle
	\begin{abstract}
		We present the first calculation of tensor and axial-tensor mesons with total spin $J\geq2$ within the Covariant Spectator Theory. We employ a refined quark-antiquark interaction kernel that incorporates the momentum dependence of the strong coupling, replacing the previously used constant term of the kernel. Global least-squares fits to the masses of experimentally established heavy and heavy-light meson states yield an excellent description of the mass spectrum for $J^P=0^\pm, 1^\pm, 2^\pm$, and $3^\pm$ using only eight adjustable parameters.
	\end{abstract}
	
\section{Introduction}
The Covariant Spectator Theory (CST) is a modern quantum-field theoretical approach for the study of few-body systems. Its simplest version is the one-channel CST, defined by the Gross equation (GE)~\cite{Gross:1969rv,PhysRevC.26.2203}. Two- and four-channel extensions of this framework were first applied to hadronic systems by Gross, Milana, and \c{S}avkl\i~in the studies of pseudoscalar and vector mesons~\cite{Gross:1991te,Gross:1991pk,Gross:1994he,Savkli:1999me}. More recently, we employed the four-channel CST to calculate the dressed quark propagator~\cite{mass_function_paper,Biernat:2018khd}, the pion form factor~\cite{pion_form_factor_paper, PhysRevD.92.076011}, and the $\pi$-$\pi$ scattering amplitude in the chiral limit~\cite{Biernat:2014xaa}. For the quark-antiquark bound state problem, we used the GE so far to compute the masses and vertex functions of heavy and heavy-light mesons with total spin $J\leq 1$~\cite{Leit_o_2017,PhysRevD.96.074007, Leitao:2017bds, Stadler:2018hjv}.
The present work extends these bound-state calculations by generalizing the formalism to mesons of arbitrary spin-parity $J^P$. This extension enables, for the first time, a unified fit of all currently established quark-antiquark meson states containing at least one heavy quark. It also predicts higher-spin states and guides the $J^P$ assignment of experimentally observed states listed by the Particle Data Group (PDG) that remain unconfirmed or have uncertain quark content. Furthermore, whereas previous CST calculations treated the strong coupling $\alpha_{\rm s}$ as a constant, we incorporate its momentum dependence here, yielding an improved overall description of the heavy and heavy-light meson spectrum.
	\section{The one-channel CST formalism}\label{sec2}
	
	The GE for a $q\bar q$ meson with spin-parity $J^P$ is given by
	\begin{equation}
		\Gamma(\hat p_1,p_2)=-\int \frac{\mathrm d^3 k_1}{(2\pi)^3}\frac{m_1}{E_{1k}} \mathcal V(\hat p_1,\hat k_1) \Lambda(\hat k_1)\Gamma(\hat k_1,k_2)  S(k_2)\, \label{eq:1CSE}
	\end{equation}
	where $\hat p_1=(E_{1p},\bm{p})$ and $\hat k_1=(E_{1k},\bm{k})$ are the external and internal on-mass-shell four-momenta of quark 1 with constituent mass $m_1$ and energies $E_{1p}=\sqrt{m_1^2+\bm{p}^2}$ and $E_{1k}=\sqrt{m_1^2+\bm{k}^2}$, respectively; the external and internal four-momenta of quark 2 are $p_2=\hat p_1-P$ and $k_2=\hat k_1-P$, respectively, where $P=(\mu,\bm{0})$ is the (on-mass-shell) four-momentum of the meson with mass $\mu$ in the rest frame. The operators
	\begin{align}
		\Lambda(\hat k_1)=\frac{m_1+\hat {\slashed {k}}_1}{2m_1}\quad
		\text{and}\quad
		S(k_2)=\frac {m_2+\slashed {k}_2}{m_2^2-k_2^2-\mathrm i \epsilon}
	\end{align}
	are, respectively, the positive-energy projector of quark 1 and the dressed propagator of quark 2 with constant constituent mass $m_2$. The CST interaction kernel is given by~\cite{PhysRevD.96.074007}
	\begin{equation}
		\mathcal V(\hat p_1,\hat k_1) = 1_1\otimes 1_2 V_{\rm L}(\hat p_1,\hat k_1) - \gamma^\mu_1\otimes\gamma_{\mu2} \left[  V_{\rm C}(\hat p_1,\hat k_1)+V_{\rm G}(\hat p_1,\hat k_1) \right]\,. \label{eq:kernel}
	\end{equation}
The three terms in Eq.~(\ref{eq:kernel}) correspond to covariant generalizations of the linear confining ($V_{\rm L}$) and a constant ($V_{\rm C}$) potentials, and a one-gluon-exchange (OGE) interaction in Feynman gauge ($V_{\rm G}$). These are given by  
	\begin{align}
	&	V_{\rm L}(\hat p_1,\hat k_1) = -8 \pi\sigma \left[ \left(\frac{1}{(\hat p_1-\hat k_1)^4}-\frac{1}{(\lambda_{\rm L}m_1)^4+(\hat p_1-\hat k_1)^4}\right) \right.\nonumber\\
		& \left. -\frac{E_{1p}}{m_1} (2\pi)^3\delta^3(\bm{p}-\bm{k}) \int \frac{\mathrm d^3 k_1'}{(2\pi)^3}\frac{m_1}{E_{1k'}}\left(\frac{1}{(\hat p_1-\hat k_1')^4}-\frac{1}{(\lambda_{\rm L}m_1)^4+(\hat p_1-\hat k_1')^4}\right)\right]\,,
	\end{align}
	\begin{equation}
	V_{\rm C}(\hat p_1,\hat k_1) = \frac{E_{1k}}{m_1} (2\pi)^3 C\delta^3(\bm{p}-\bm{k}) \,,
	\end{equation}
	and
	\begin{equation}
		V_{\rm G}(\hat p_1,\hat k_1) = -\frac{16\pi }{3}\alpha_{\rm s} \left((\hat p_1-\hat k_1)^2\right) \left(\frac{1}{(\hat p_1-\hat k_1)^2}-\frac{1}{(\hat p_1-\hat k_1)^2-(\lambda_{\rm G}m_1)^2}\right)\,,
	\end{equation} 
where the strong running coupling is
	\begin{equation}
	\alpha_{\rm s}(q^2) = \frac{1}{\beta_0 \ln\left(\frac{-q^2}{\Lambda_{\rm QCD}}+\tau\right)} \, ,
	\end{equation}
	with $\beta_0=\frac{33-2N_f}{12\pi}$, and $N_f$ is the number of active flavours. The IR regulator $\tau$ is fixed by specifying the value of $\alpha_s(q^2)$ at $q^2=0$. The remaining parameter, $\Lambda_{\rm QCD}$, is determined by requiring that $\alpha_s(q^2)$ reproduces the experimental value at $q^2=-M_Z^2$. We regularize the UV-behaviour of linear-confining and OGE kernels via Pauli-Villars subtraction, leaving the corresponding dimensionless cut-off parameters ($\lambda_{\rm L}$ and $\lambda_{\rm G}$) and the coupling strengths ($\sigma$, $\alpha_{\rm s}(q^2)|_{q^2=0}$, and $C$) as adjustable fit parameters.  
	
	The spin-$J$ meson vertex function $\Gamma(\hat p_1,p_2)$ can be written, depending on the parity $P$, as
	\begin{align}
		&\Gamma(\hat p_1,p_2) = \zeta^{\mu\nu\sigma\ldots}_{m_J} \nonumber\\
		&\times \begin{cases} 
			F M_{\mu\nu\sigma\ldots}  +GN_{\mu\nu\sigma\ldots} + \left[HM_{\mu\nu\sigma\ldots}  +IN_{\mu\nu\sigma\ldots} \right]\Lambda (-p_2),& P=(-1)^J \\
			\left\{ \tilde F M_{\mu\nu\sigma\ldots}+\tilde GN_{\mu\nu\sigma\ldots} + \left[\tilde HM_{\mu\nu\sigma\ldots} +\tilde IN_{\mu\nu\sigma\ldots}\right]\Lambda (p_2)\right\}\gamma^5 ,& P=(-1)^{J+1} 
		\end{cases}
	\end{align} 
	where
	\begin{equation}
		\zeta^{\mu\nu\sigma\ldots\omega}_{m_J} = \left\lbrace \xi_{\lambda_1}^\mu\otimes \xi_{\lambda_2}^\nu\otimes\cdots\otimes \xi_{\lambda_J}^\omega \right\rbrace_{m_J}^{(J)}
	\end{equation}
	is the rank-$J$ spherical polarization tensor associated with angular momentum $J$ and spin-polarization $m_J=-J,\ldots,J$. The four-vectors $\xi_{\lambda_i}^\mu$ are the three spherical polarization vectors for massive spin-1 states, corresponding to spin polarizations $\lambda_i=\pm 1$ and $0$. The shorthand functions $F, G, \ldots,   \tilde I$ implicitly depend on the Lorentz invariants $\hat p_1\cdot p_2$ and $p_2^2$, and 
	\begin{align}
		M^{\mu\nu\sigma\ldots\omega} &= \begin{cases}
			\frac{1}{J}\left(\gamma^\mu p^\nu p^\sigma\cdots p^\omega+p^\mu\gamma^\nu  p^\sigma\cdots p^\omega +\ldots+p^\mu p^\nu p^\sigma  \cdots \gamma^\omega\right), & J>0 \\
			0, & J=0
		\end{cases} \nonumber\\
		N^{\mu\nu\sigma\ldots\omega} &= \begin{cases} 
			p^\mu p^\nu p^\sigma \cdots p^\omega, & J>0 \\
			\mathbf{1}, & J=0 
		\end{cases} 
	\end{align}  
	are rank-$J$ Lorentz tensors.
	
For the identification of states listed by the PDG, it is convenient to switch from the Lorentz-tensor basis to a partial-wave basis by decomposing Eq.~(\ref{eq:1CSE}) into its positive- and negative-energy channels. The resulting vertex-function matrix elements between $u$ and $v$ Dirac spinors can be expanded for a fixed $J^P$ in terms of orbital angular momentum $L$ and spin $S$ eigenfunctions. The resulting eigenvalue problem is then solved for the set of possible partial waves and the corresponding energy eigenvalue $\mu$.
	
\section{Results and Conclusions}

We have solved the GE, Eq.~(\ref{eq:1CSE}), for each meson sector from $b\bar b$ to $c\bar q$ (where $q=u, d$), and the $J^P=0^\pm, 1^\pm, 2^\pm, 3^\pm$ channels in each sector. The initially nine model parameters (including the constant $C$ and the constituent quark masses $m_u, m_s, m_c$, and $m_b$) were adjusted through global least-squares fits to experimentally measured states. We evaluated three parameter models, differing by the set of states included in the fit: 10 pseudoscalar states, 33 non-axial states ($J^\mathcal P=0^\pm, 1^-, 2^+, 3^-$), and 49 states of all channels. 

The parameter values for the three models are listed in Table~\ref{tab1}. Note that the strength $C$ of the constant interaction is omitted from the table because the fit consistently yields $C\approx 0$. In models with a fixed $\alpha_{\mathrm s}$, the constant potential effectively simulates the missing running behaviour. By explicitly incorporating the momentum dependence of $\alpha_{\mathrm s}$, the constant interaction becomes redundant, leaving us with only eight adjustable model parameters. Furthermore, implementing the running of $\alpha_{\mathrm s}$ substantially improves the global fit compared to using a constant $\alpha_{\mathrm s}$. The best overall agreement is obtained with $N_f=2$ active flavours, which performs slightly better than $N_f=3$.  
	
\begin{figure}[htb]
	\centerline{%
		\includegraphics[width=0.5\textwidth ]{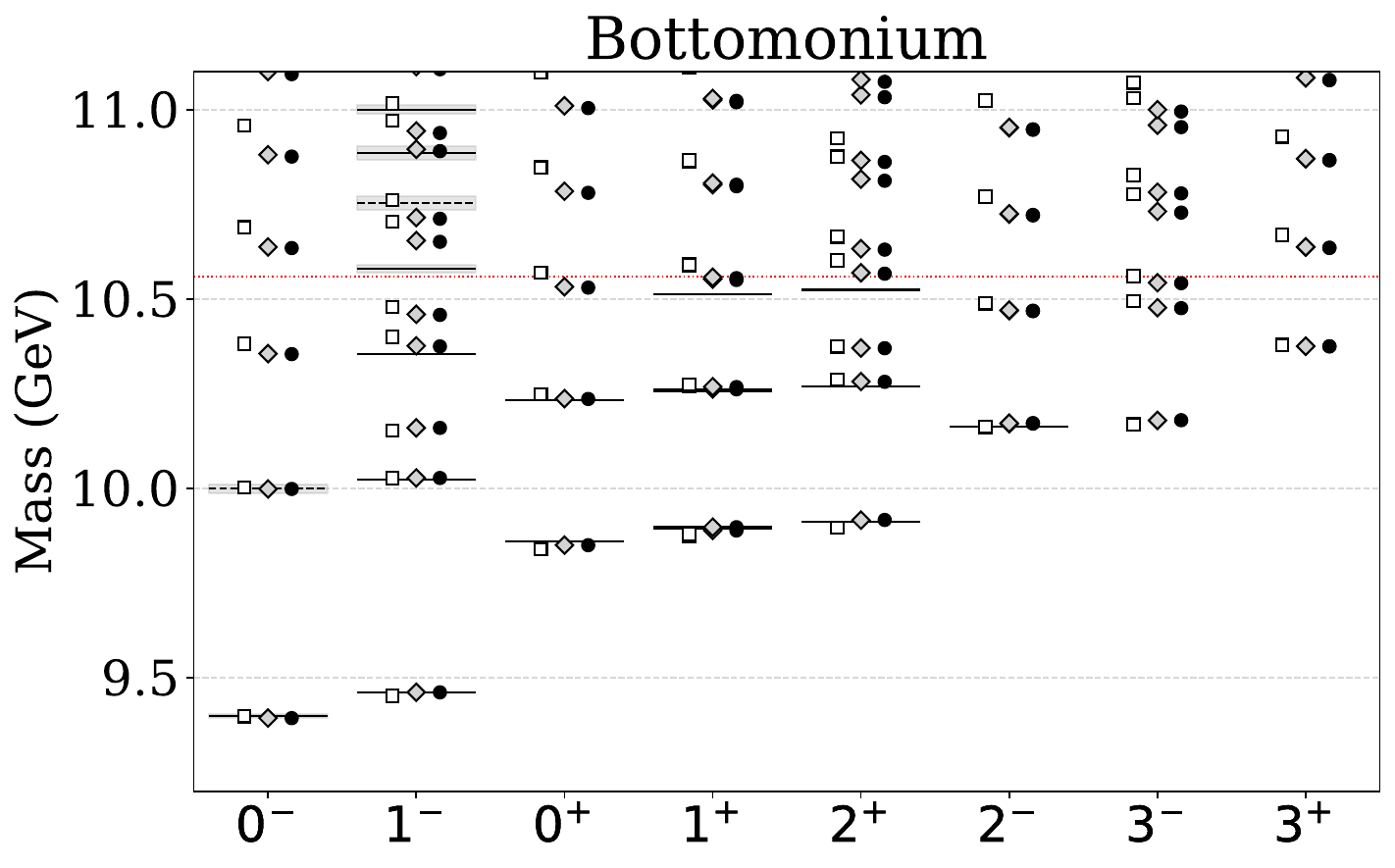}\includegraphics[width=0.5\textwidth]{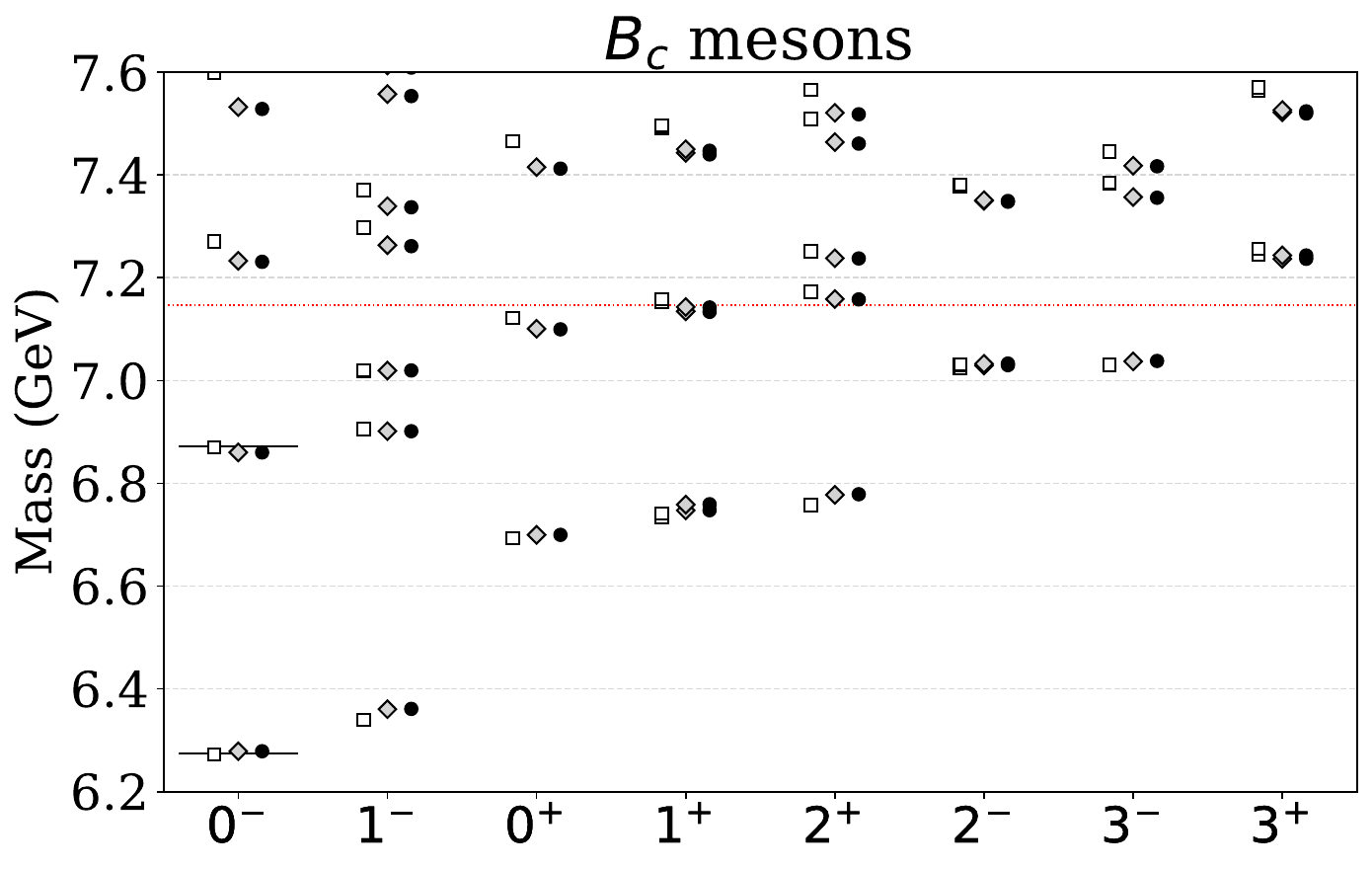}}
	\centerline{\includegraphics[width=0.5\textwidth ]{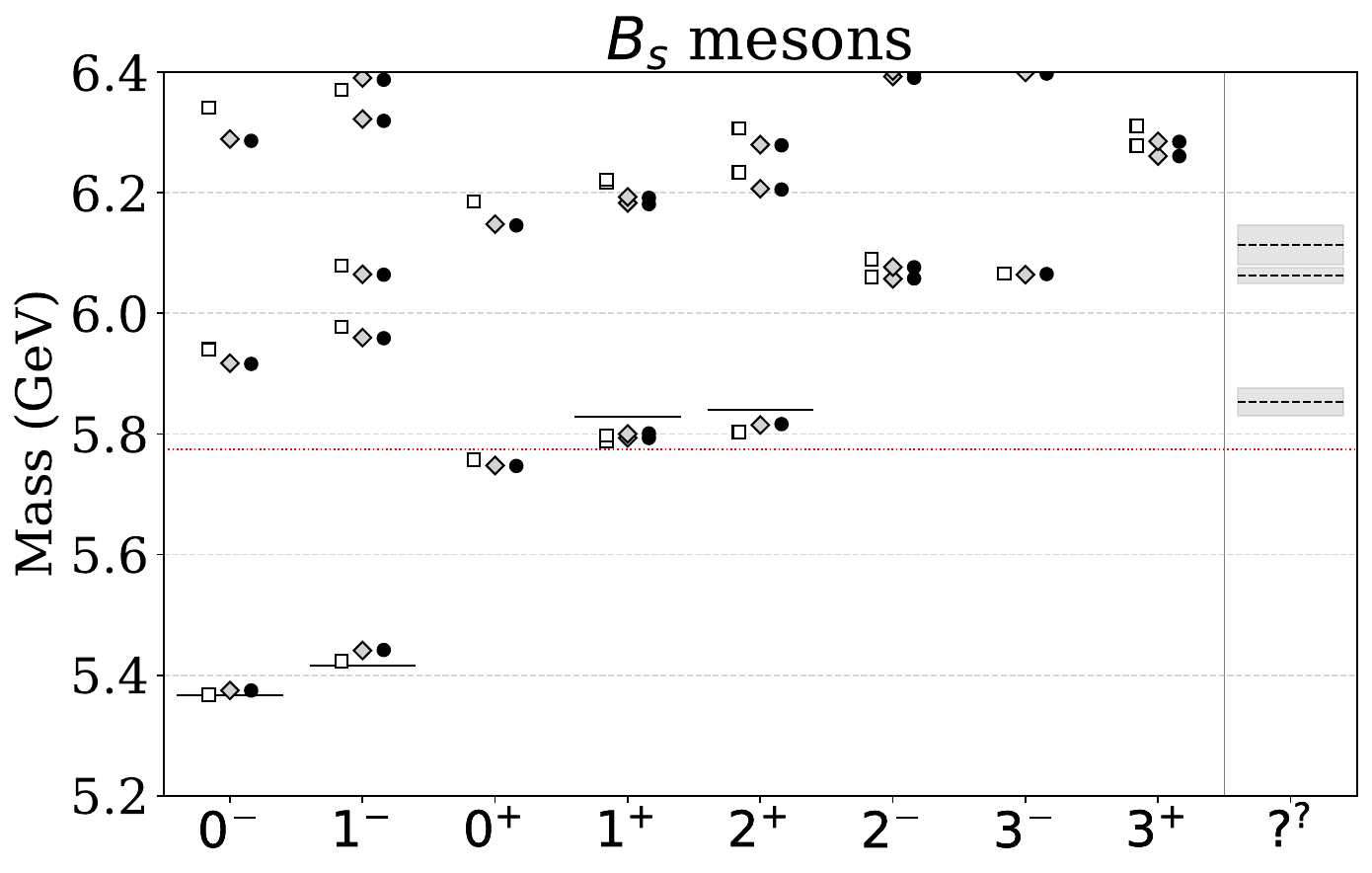}\includegraphics[width=0.5\textwidth]{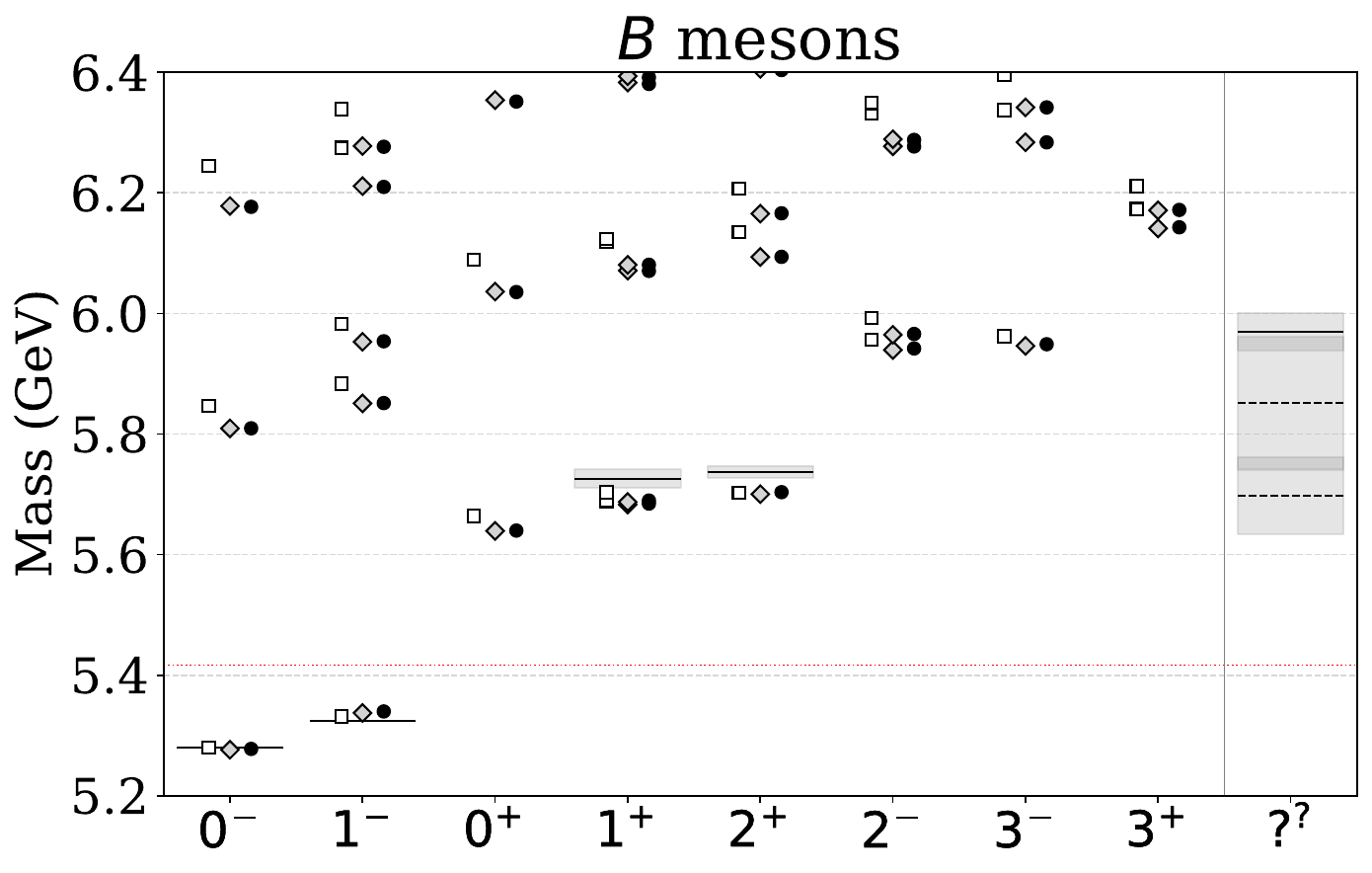}}
	\centerline{\includegraphics[width=0.5\textwidth ]{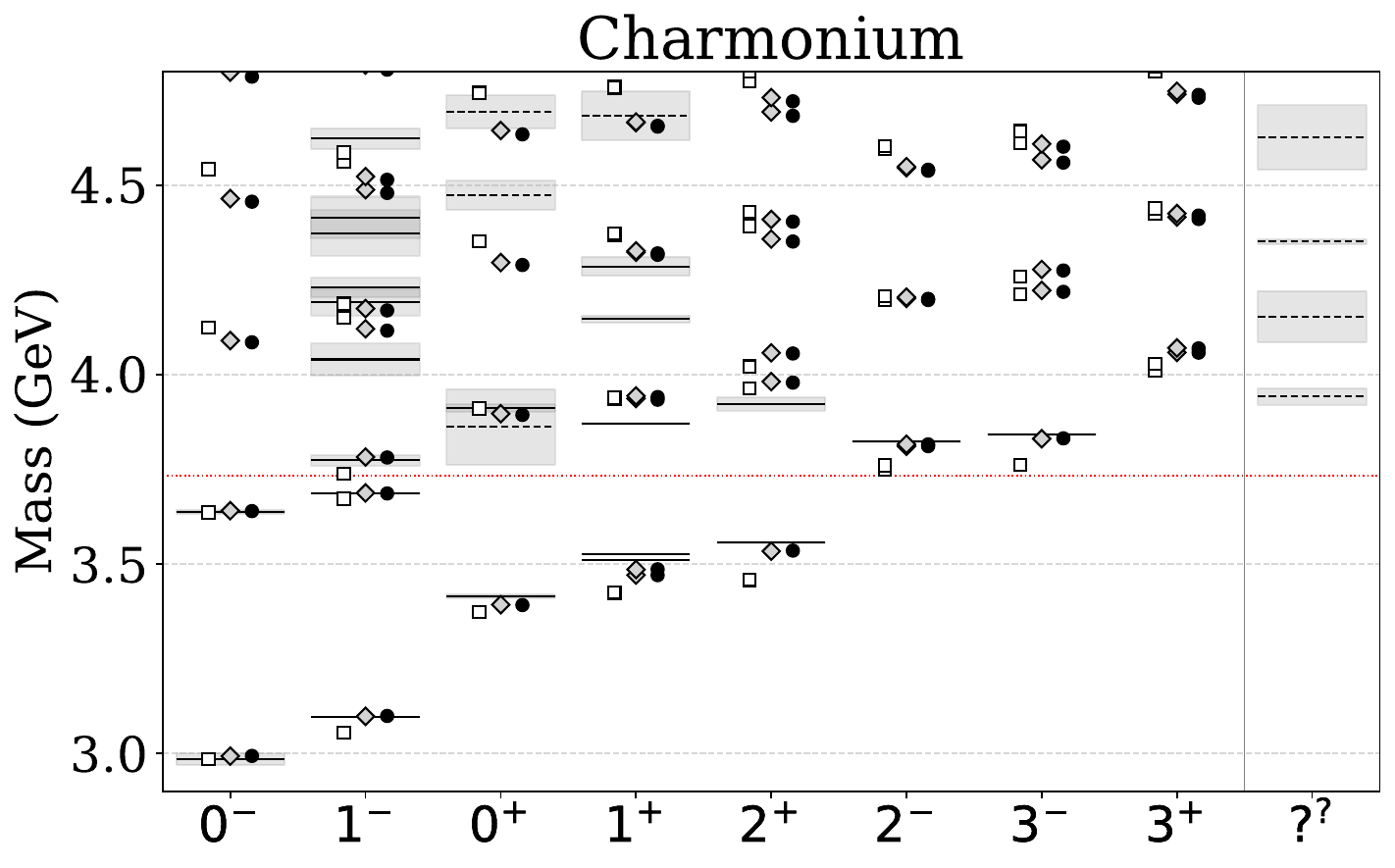}\includegraphics[width=0.5\textwidth]{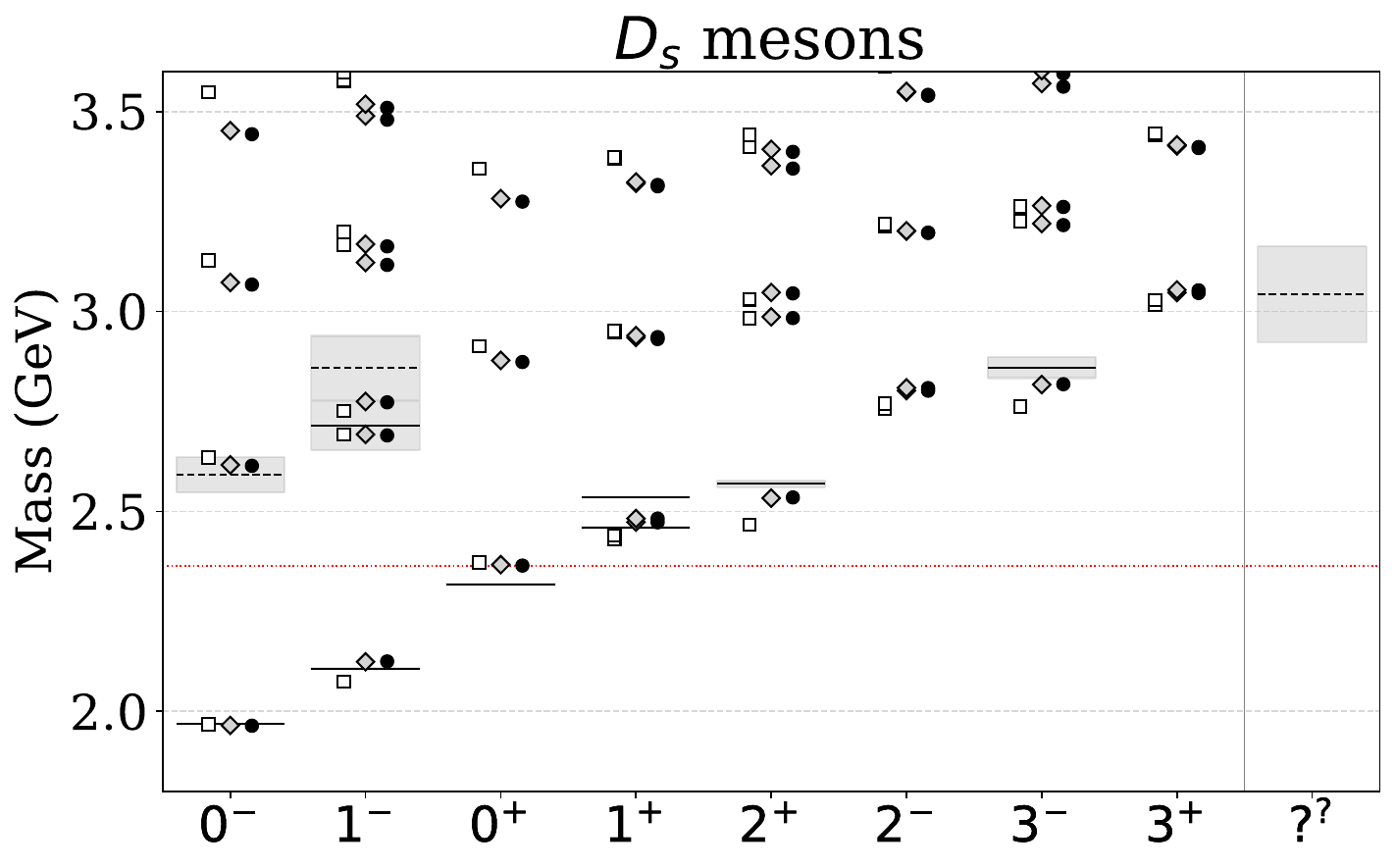}}
	\centerline{%
		\includegraphics[width=0.5\textwidth ]{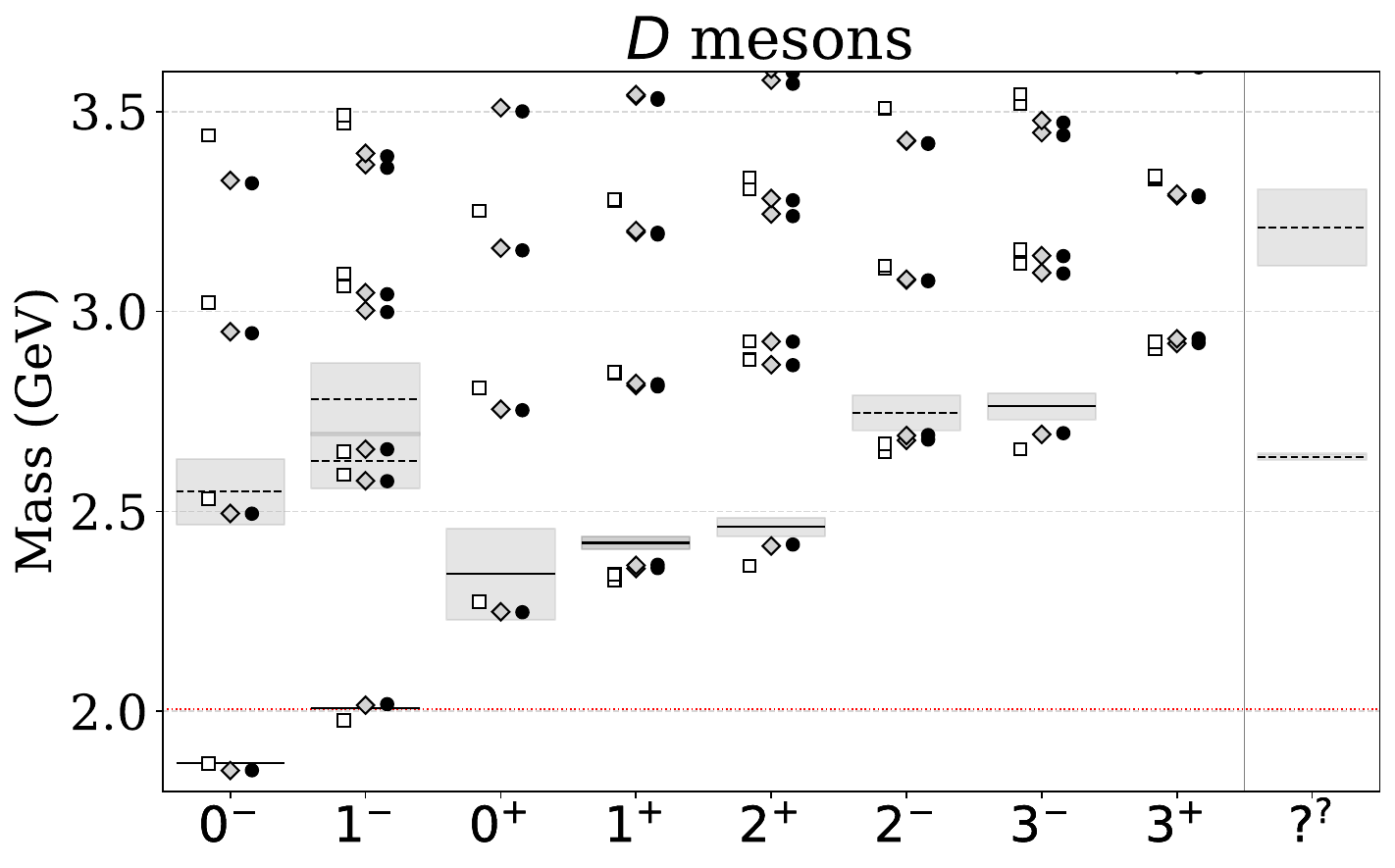}}
	\caption{The results of fits to pseudoscalar (empty squares), non-axial (grey-filled diamonds), and  all established (black dots) states, compared with the established (solid lines) and unconfirmed (dashed lines) experimental data, with the grey shading displaying the width. The red dotted lines indicate the open-flavour thresholds.}
	\label{fig1}
\end{figure}

\begin{table}[htb]
	\centering
	\caption{Parameter values of the three fit models with $N_f=2$.}
	\label{tab1}
	\begin{tabular}{l|cc|cccc|cc}
		\hline
		\hline
		Fitted   & \multicolumn{2}{|c|}{Strengths} & \multicolumn{4}{c}{Masses [GeV]}&\multicolumn{2}{|c}{Cut-offs} \\
	states& $\sigma$ [GeV$^2$]& $\alpha_{\rm s}(0)$&$m_b $&$m_c $&$m_s $&$m_q $&$\lambda_{\rm L}$&$\lambda_{\rm G}$ \\
		\hline
		10 & 0.216 & 0.419 & 4.794 & 1.441 & 0.274 & 0.133 & 1.219 & 1.786 \\
		33 & 0.179 & 0.507 & 4.852 & 1.508 & 0.343 & 0.185 & 2.812 & 2.266 \\
		49 & 0.176 & 0.522 & 4.859 & 1.517 & 0.353 & 0.197 & 2.903 & 2.243 \\
		\hline
		\hline
	\end{tabular}
\end{table}
The resulting mass spectra for the three models are presented in Fig.~\ref{fig1}.
We find that fitting only the pseudoscalar states yields fairly accurate predictions for the remaining $J^P$ channels, including the higher-spin tensor states. The most significant deviations from the other two models occur primarily in the higher excited states. As expected, fitting all established states provides the best overall description of the spectrum. In fact, it yields results very similar to fitting only the non-axial states (for which the GE is best suited), a fact explicitly reflected in the similar parameter values shown in Table~\ref{tab1}.

Our results confirm that earlier conclusions~\cite{Leit_o_2017,PhysRevD.96.074007} extend to tensor mesons: the covariant structure of the kernel correctly captures the spin dependence of the $q\bar q$ interaction, which allows for predictions of higher-spin states. This provides a unified description of the mass spectrum for $q \bar q$ mesons containing at least one heavy quark. Ultimately, this work completes our treatment of these states using the GE and establishes groundwork for modeling light mesons of arbitrary $J^P$ via the four-channel CST equation.

\subsection*{Acknowledgments}
This work was supported by FCT under the project reference UID/04349/2025 (DOI: https://doi.org/10.54499/UID/04349/2025). 
\bibliographystyle{apsrev}
\bibliographystyle{unsrt}
\bibliography{PapersDB}
\end{document}